\documentclass[12pt]{article}
\usepackage{graphicx,amssymb}
\usepackage{qcmc06}
\def\Tr{\hbox{Tr}}

\begin{document}
\pagestyle{empty}
\title{Information-disturbance tradeoff 
in covariant \\ 
quantum state estimation}
\author{Massimiliano F. Sacchi}

\affiliation{
CNISM and CNR - Istituto Nazionale per la Fisica della Materia,\\
  Dipartimento di Fisica ``A. Volta'',  via A. Bassi 6, I-27100 Pavia, Italy} 

\begin{abstract} We provide a general framework for quantifying the
optimal tradeoff between the information retrieved by a quantum
measurement and the disturbance on the quantum state in covariant
quantum state estimation.  
\end{abstract} \setcounter{section}{0}


There exists a precise tradeoff between the amount of
information extracted from a quantum measurement and the amount of
disturbance caused on the system, analogous to Heisenberg relations
holding in the preparation procedure of a quantum state. 
The study of such a tradeoff is relevant 
for both foundation and its enormous relevance in practice,
in the realm of quantum key distribution and quantum cryptography. 
Quantitative derivations of such a tradeoff have been obtained in
the scenario of quantum state estimation, in estimating a
single copy of a pure state \cite{banaszek01.prl}, many
copies of identically prepared pure qubits \cite{banaszek01.pra}, a
single copy of a pure state generated by independent phase-shifts
\cite{mista05.pra}, an unknown maximally entangled state \cite{max},  coherent state 
\cite{cv}, and spin-coherent state \cite{scs}, 
and in state discrimination of two pure states \cite{sd}.  

In this paper, we provide a unified framework to study 
the optimal tradeoff between information and disturbance 
for set of states with given symmetry. 
Our results will be obtained by exploiting the group symmetry of the
problem, which allows us to restrict our analysis on {\em covariant
  measurement instruments}. In fact, the property of covariance
  generally leads to a striking simplification of problems that may
  look intractable, and has been thoroughly used in the context of
  state and parameter estimation. 

The problem of the tradeoff in covariant state estimation is the following. 
One performs a measurement on
a quantum state picked randomly from a known set, and evaluates the retrieved
information along with the disturbance caused on the state. To
quantify the tradeoff between information and disturbance, one can
adopt two mean fidelities \cite{banaszek01.prl}: the estimation
fidelity $G$, which evaluates on average the best guess we can do of
the original state on the basis of the measurement outcome, and the
operation fidelity $F$, which measures the average resemblance of the
state of the system after the measurement to the original one. The solution of the 
optimal tradeoff provides a set of minimum-disturbing measurements, 
that for fixed value of $G$ maximizes $F$. 

A measurement process on a quantum state $\rho $ with outcomes $\{r
\}$ is described by an {\em instrument},  
namely a set of
trace-decreasing completely positive (CP) maps $\{{\cal E}_r \}$. Each
map can then be written in the Kraus form 
\begin{eqnarray}
{\cal E}_r( \rho )= \sum _\mu  A_{r \mu} \rho A_{r \mu} ^\dag \;,
\label{uno}
\end{eqnarray}
and provides the state after the measurement 
$\rho _r =\frac{{\cal E}_r
(\rho )}{\Tr [{\cal E}_r (\rho )]}$, 
along with the probability of
outcome 
$p_r = \Tr [{\cal E}_r (\rho )] = \Tr\left [\sum _\mu A^\dag _{r
  \mu}A_{r\mu } \rho\right ]$.

The set of positive operators $\{ \Pi _r = \sum _\mu
A^\dag _{r \mu}A_{r\mu }\}$ is known as positive operator-valued
measure (POVM), and normalization requires the completeness relation
$\sum _r \Pi _r = I$. This is equivalent to require that the map $\sum
_r {\cal E}_r $ is trace-preserving.

We are interested in covariant state estimation, a problem where we want to estimate 
a quantum state that belongs to a covariant set of states 
\begin{equation}
\{ |\psi  _g \rangle = U_g |\psi _0 \rangle \} \,,
\end{equation} 
where $|\psi _0 \rangle$ is a fixed reference state of a Hilbert space $\cal H$ 
with finite dimension 
$\hbox{dim}({\cal H})=d$, and $U_g $, $g \in {\bf G}$,  
is unitary representation of a group $\bf G$ on $\cal H$.  
We will consider
only compact groups (which then admit normalizable invariant Haar
measure $dg$), and for convenience we will take the normalized
invariant Haar measure over the group, i.e. $\int_{\bf G} dg=1$. 
The unknown state is then randomly distributed  according to $dg$. 

\par The operation fidelity $F$ evaluates on average how much the
state after the measurement resembles the original one, in terms of
the squared modulus of the scalar product.  Hence, for a measurement described by 
(\ref{uno}) one has
\begin{eqnarray}
F = \int dg \sum _{r \mu}  
|\langle \psi _g  |A_{r \mu } |\psi_g \rangle |^2 \;.\label{fm}
\end{eqnarray}
By adopting a guess function $f$, 
for each measurement outcome $r$ one guesses
a spin coherent states $|\psi _ {f(r)} \rangle$, and the
corresponding average estimation fidelity is given by 
\begin{eqnarray}
G
=
\int  dg \sum _{r \mu}  \langle \psi_g | A^\dag _{r\mu}
A_{r \mu}|\psi _g \rangle   
\,|\langle \psi_{f(r)}|\psi_g \rangle |^2 
\;.\label{gm}
\end{eqnarray}
We are interested in the optimal tradeoff between $F$ and $G$, and 
without loss of generality we can restrict our attention to {\em
  covariant} instruments, that satisfy 
\begin{eqnarray}
{\cal E} _h (U_g 
\,\rho \, U^\dag _g) 
=U_g {\cal E}_{g^{-1}h}(\rho )U^\dag _g 
\;.\label{cvv}
\end{eqnarray}
In fact, 
for any instrument $\{ A_{r \mu }\}$ and guess function $f$ 
the covariant instrument
\begin{eqnarray}
{\cal E}_h (\rho )  = \sum _{r \mu} 
U_h U^\dag _{f(r)}  A_{r \mu} U_{f(r)} 
U_h^\dag 
  \,\rho \, U_h U^\dag _{f(r)}  A_{r \mu} ^\dag U_{f(r)} 
  U_h^\dag  
\end{eqnarray} 
with continuous outcome $h\in {\bf G}$,  
along with the guess $|\psi _h \rangle $, provides the
same values of $F$ and $G$ as the original instrument. 
Moreover, for covariant instruments the optimal guess function
automatically turns out to be the identity function.

It is useful now to consider the Jamio\l kowski representation 
that gives a one-to-one correspondence between a CP map ${\cal E}$
from ${\cal H}_{in}$ to ${\cal H}_{out}$ and a
positive operator $R$ on ${\cal H}_{in}\otimes {\cal H}_{out}$ 
through the equations 
\begin{eqnarray}
&&{\cal E}(\rho )=\Tr _{in}[(\rho ^\tau \otimes I_{out} ) R ]\;,
\nonumber \\& & 
R=(
I_{in} \otimes {\cal E }
) |\Phi \rangle \langle \Phi | \;,
\label{jam}
\end{eqnarray}
where $|\Phi \rangle = \sum _{i =1}^d |i \rangle \otimes |i\rangle $  
represents the unnormalized maximally entangled vector of 
${\cal H}_{in}^{\otimes 2}$, 
and $\tau $ denotes the transposition on the fixed basis.  
When ${\cal E}$ is trace preserving, correspondingly one has $\Tr_{out}[R] = I_{in}$. 

For covariant instruments ${\cal E}_g$  the
operator $R_g$ has the form
\begin{eqnarray} 
R_g = U_g ^{*}  \otimes U_g 
\,R_0 \, U_g ^{\tau  }\otimes
U_g ^{\dag } \;,\label{rggg} 
\end{eqnarray} 
where $*$ denotes the complex conjugation, and the trace-preserving condition is given by 
\begin{eqnarray} \int dg \, \Tr _{out}[R_g] =
\int dg \, U^* _g \,\Tr _{out}[R_0] \,U^\tau _g  =
I _{in}\;.\label{rgg}
\end{eqnarray}
The fidelities  
$F$ and $G$ in Eqs. (\ref{fm}) and (\ref{gm}) can be rewritten as follows
\begin{eqnarray}
F&=& \int dg \int dh \, 
\langle \psi _g |{\cal E}_h (|\psi _g \rangle \langle \psi _g |)|\psi 
_g \rangle = \int dg \, 
\langle \psi _0 
|{\cal E}_g (|  \psi _0 \rangle \langle \psi _0  |)| \psi _0  \rangle\;,\\
G&=& \int dg \int dh \, |\langle \psi _g |\psi _h \rangle |^2 \,
\Tr [{\cal E}_h (|\psi _g \rangle \langle \psi _g |)] =
 \int dg \,|\langle \psi _0  |U_g | \psi _0  \rangle |^2
\Tr [{\cal E}_g (| \psi _0  \rangle \langle  \psi _0 |)]\;,
\end{eqnarray}
where the covariance property (\ref{cvv}) and the invariance of the 
Haar measure have been used. Using the isomorphism (\ref{jam}), 
we can write $F$ and $G$ as 
$F=\Tr [R_F R_0]$ and $G=\Tr [R_G R_0]$, where $R_F$ and
$R_G$ are the following positive operators 
\begin{eqnarray} R_F &=&
\int dg \,|\psi _g \rangle \langle \psi _g|
^\tau \otimes 
|\psi _g \rangle \langle \psi _g |\;,\\
R_G&=& 
\Tr _{out} [(I_{in} \otimes 
|\psi _0 \rangle \langle \psi _0 |)R_F]\otimes I_{out}\;.
\end{eqnarray}

The optimal tradeoff between $F$ and $G$ can be found by 
maximizing the operation fidelity
$F=\Tr[R_F R_0]$ versus $R_0$, for  fixed value of the estimation fidelity
$G=\Tr[R_G R_0]$ and under the constraint (\ref{rgg}).

The evaluation of the constraint (\ref{rgg}) and 
the operator $R_F$ (and hence of $R_G$) needs 
group averages with the representations $V_g =U^* _g$ and $V_g=U_g ^*\otimes U_g $, 
respectively.   These can
be obtained as follows.  The Hilbert space can be decomposed into
orthogonal subspaces \begin{equation}\label{SpaceDecomp} {\cal H}
\equiv \bigoplus_{\mu \in S}~ {\cal H}_{\mu} \otimes \mathbb C
^{m_{\mu}}~, \end{equation} where the sum runs over the set $S$ of
irreducible representations that appear in the Clebsch-Gordan
decomposition of $V_g $.  The action of the group is
irreducible in each representation space ${\cal H}_{\mu}$, while it is
trivial in the multiplicity space ${\mathbb C}^{m_{\mu}}$, which takes
into account the presence of equivalent representations.

From Schur's lemma one has 
\begin{eqnarray} 
\int dg \,V_g \,Y \, V_g ^\dag  
= \sum_{\mu } I_{d_\mu }\otimes \frac{\Tr _{{\cal H}_\mu }[Y P_{\mu }]}{d_\mu }
\;,
\end{eqnarray}
where $\{ P_\mu \}$ are the orthogonal projectors 
on the invariant subspaces ${\cal H}_{\mu} \otimes {\mathbb C}^{m_{\mu}}$, and 
$I_{d_\mu }$ denotes the identity operator on ${\cal H}_\mu $ (which has dimension 
$d_\mu $).  

When the irreducible representations of $V_g $ are all
inequivalent, notice that all $m_\mu $ are equal to one.  This is the
case of all examples given in the following.

$i)$ For unknown pure state with $\hbox{dim}({\cal H})=d$, one has $|\psi _g \rangle 
= U_g |\psi _0 \rangle $, with $|\psi _0 \rangle $ arbitrary, $U_g \in SU(d)$ and 

\begin{eqnarray}
R_F 
&= & \frac{1}{d(d+1)}
( I+ {\cal I}) \;,\\
R_G&=& \frac{1}{d(d+1)}
(I+ |\psi _0\rangle \langle \psi _0 | ^\tau )\otimes I \;,
\end{eqnarray}
where ${\cal I}=(\sum _{n=1}^{d}
|n\rangle \otimes |n \rangle ) \,(\sum _{m=1}^{d}
 \langle m | \otimes  \langle m|)$. 

$ii)$ For an unknown  maximally entangled state of ${\cal H}^{\otimes 2}$ with 
$\hbox{dim}({\cal H})=d$, one has 
$ \{ |\psi  _g \rangle = (U_g \otimes I )|\psi _0 \rangle \} $ with 
$ |\psi _0 \rangle = \frac 1{\sqrt d}\sum _{n=1}^{d} 
|n \rangle \otimes |n \rangle $, $U_g \in SU(d)$ and \cite{max}

\begin{eqnarray}
R_F 
&= & 
\frac{1}{d^2 (d^2 -1)} \left [ I + {\cal I}^{(13)}\otimes
{\cal I}^{(24)}- 
\frac 1d (I^{(13)}\otimes {\cal I}^{(24)}
+{\cal I}^{(13)}\otimes I^{(24)}
)\right ]
\;,
\\
R_G &= &  
\frac{1}{d^2 (d^2 -1)} \left [\left (1 -\frac{2}{d^2}\right )I + \frac 
1d I^{(12)}\otimes
{\cal I}^{(34)}\right ] \;.
\label{rgg2}
\end{eqnarray}
where ${\cal I}^{(ij)}=(\sum _{n=1}^{d}
|n\rangle _i \otimes |n \rangle _j )
(\sum _{m=1}^{d} {}_i \langle m | \otimes {}_j \langle m|)
$.

$iii)$ For spin-$j$ coherent states, $ \{ |\psi  _g \rangle = U_g |-j \rangle \} $, where 
$|-j \rangle $ is the eigenvector of $J_z$ with minimum eigenvalue, 
$ U_g$ is a  $(2j+1)$-dimensional irreducible representation of $SU(2)$, and 
\cite{scs}
\begin{eqnarray}
R_F 
&= & 
\frac{1}{4j+1} P_{2j}^\theta\;,
\\
R_G &= &  
\frac{1}{4j+1} 
\Tr _2 [(I \otimes |-j \rangle \langle -j | ) P_{2j}]
\otimes I 
\;,
\label{rgg3}
\end{eqnarray}
where $\theta $
denotes the partial transpose on the first Hilbert space, and $P_{l}$
represents the projector on the subspace of ${\cal H}\otimes {\cal H}$
with total spin $l$.

\ack

This work has been sponsored by Ministero 
Italiano dell'Universit\`a e della Ricerca (MIUR) through FIRB (2001)
and PRIN 2005.

\end{document}